\documentclass{article}
\newtheorem{theorem}{Theorem}

\begin{document}
%\draft

% remove the following line for the preprint format
%\twocolumn[\hsize\textwidth\columnwidth\hsize\csname@twocolumnfalse\endcsname

\title{
Information-Disturbance theorem
 and Uncertainty Relation
}

\author{Takayuki Miyadera$\ ^*$
%\cite{MIYA} 
and Hideki Imai$\ ^{*,\dagger}$
%}
%
%\address{
%\begin{center}
\\
$\ ^*$
Research Center for Information Security (RCIS), \\
National Institute of Advanced Industrial \\ 
Science and Technology (AIST). \\
Daibiru building 1102,\\
Sotokanda, Chiyoda-ku, Tokyo, 101-0021, Japan.
\\
(e-mail: miyadera-takayuki@aist.go.jp)
\\
$\ ^{\dagger}$
Graduate School of Science and Engineering,
\\
Chuo University. \\
1-13-27 Kasuga, Bunkyo-ku, Tokyo 112-8551, Japan .
}
%\end{center}
%\abstract{
%}
% \date{Compiled: \today}

\maketitle
\begin{abstract}
It has been shown that Information-Disturbance theorem 
can play an important role in security proof of quantum cryptography.
The theorem is by itself interesting since it can be 
regarded as an information theoretic version of uncertainty 
principle. 
%We, in previous papers, have shown a novel type of 
%the theorem that relates information gain by Eve and 
%randomness of error contained in Bob's outcome. 
It, however, has been able to treat restricted situations. 
%and needs to be generalized for application to more realistic 
%circumstances. 
In this paper, %We, in this paper, get rid of the 
the restriction on the source %that was imposed, and derive a rather 
is abandoned, and 
a general information-disturbance theorem 
is obtained. %Our new theorem 
%The new theorem can be applied to a quantum cryptographic setting with 
%arbitrary source.  
The theorem relates information gain by Eve 
with information gain by Bob. 
\end{abstract}
%\pacs{PACS numbers:  03.65.Ta, 03.67.-a}
%]

\section{Introduction}
In 1984, Bennett and Brassard\cite{BB84} proposed a 
protocol to realize key distribution
that uses quantum theory in its essential part. 
In spite of simplicity of the protocol, its 
unconditional security proof\cite{Mayers,HKL,Shor,Biham} appeared 
more than a decade later after its proposal. 
Among the various existing proofs, 
a proof by Biham et al.\cite{Biham} employs a so-called 
information-disturbance theorem\cite{FuchsPeres,Fuchs,Winter,Boykin} 
that can be 
regarded as an information theoretical version 
of the uncertainty relation. 
We, in \cite{della}, succeeded in deriving an improved
variation of the theorem. Our theorem expressed a relation between 
information gain by Eve and randomness of error contained in Bob's data.
Although it has a natural form, 
its applicability is still restricted. 
In fact the state prepared by Alice has to be 
ensembles consisting of pure states with even probability. 
%On the other hand, 
%when we think of the real experiment, this condition 
%is not satisfactory sufficiently. 
%In fact it is difficult to prepare (pure) states exactly and 
%to measure some observables without error.
In this paper, we get rid of this strong condition and show 
fairly generalized form of the information-disturbance theorem. 
Alice prepares an arbitrary state by one of two different 
ensembles.  
That is, Alice chooses one of two random variables to 
be encoded. Each ensemble does not need to consist of 
distinguishable states.  
Our new information-disturbance theorem represents 
a relation between Eve's information gain and 
Bob's information gain. 
According to the theorem, if Eve employs an attack that gives 
her large information on 
an encoded random variable, Bob could obtain small information 
on another random variable.
This trade-off is determined by noncommutativity between 
the ensembles. 
The theorem is derived by using 
remote ensemble preparation technique and entropic uncertainty 
relation. 
These technique also allows us to obtain 
a simple derivation of the result in \cite{della}. 
In section\ref{preliminaries}, we give a brief review on 
positive operator valued measure and entropic uncertainty relation that 
play central roles in our proof. In section\ref{rep}, 
we introduce a method to prepare remotely an ensemble of 
quantum states by making a proper measurement on 
predistributed quantum state. 
In section\ref{infodis}, our main theorems are presented. 
%%%%%%%%%%%%%%%%
\section{Preliminaries}\label{preliminaries}
We begin with a brief introduction of relevant notions
in quantum theory: positive operator valued measure and 
entropic uncertainty relation.
%%%%%%%%%%%%%%%%%%%%%%%%%%%%%%%%%%%%%
\subsection{Positive Operator Valued Measure (POVM)}
A quantum system is described by a Hilbert space and 
operators acting on it. 
%(The most general setting
%that allows us to treat infinite systems consists of 
%an abstract operator algebra and opertation on them, we, however, 
%do not treat it here.) 
The most general observable is represented by a 
positive operator valued measure (POVM) (see, e.g.
\cite{NC}). 
A positive operator valued measure $A(\cdot)$ is 
a map from measurable space $(\Omega,{\cal F})$ 
to a set of 
positive operators satisfying:
\begin{itemize}
\item[(i)]
For all $S_1,S_2\in {\cal F}$ 
satisfying $S_1\cap S_2 =\phi$, 
$A(S_1 \cup S_2)=A(S_1)+A(S_2)$
holds. 
\item[(ii)]
$A(\Omega)={\bf 1}$ holds.
\end{itemize}
Hereafter we treat only the case that 
the measurable set is a finite set. Therefore 
the conditions above can be rephrased as 
follows. 
A POVM is a family of positive operators 
$\{A_a\}_{a\in \Omega}$ satisfying 
$\sum_{a\in \Omega} A_a ={\bf 1}$. 
Each $a\in \Omega$ corresponds to a measurement outcome. 
A POVM is called as a projection valued measure (PVM) 
if $A_a$ is a projection operator for all $a \in \Omega$.
A state is described by a so-called density operator. 
A density operator $\rho$ is 
defined by an operator satisfying  
$\rho\geq 0$ and $\mbox{tr}\rho=1$. 
If one measures an observable $A=\{A_a\}$ 
in a state $\rho$, one obtains an outcome 
$a$ with probability $\mbox{tr}(\rho A_a)$.
From a POVM $A=\{A_a\}$ one can construct 
a self adjoint operator $\hat{A}:=\sum_{a\in \Omega}
a A_a$. This operator is useful since 
it gives the expectation value for the 
measurements so that $\langle A\rangle_{\rho}
=\mbox{tr}(\rho \hat{A})$. 
For PVM, the standard deviation can be calculated 
as $\Delta A_{\rho}
=
(\langle \hat{A}^2\rangle_{\rho}-\langle \hat{A}\rangle^2_{\rho})^{1/2}$.
%%%%%%%%%%%%%%%%%%
\subsection{Entropic Uncertainty Relation}
As is widely known, the uncertainty relation is the most fundamental 
result of quantum theory. It, in general, is expressed by an 
inequality. The uncertainty relation treats two (or more) observables. 
Incompatibility of probability distributions 
of their measurement 
outcomes is bounded by noncommutativity between them. 
The most famous one is the Robertson-type uncertainty relation
for PVMs:
%\\
%{\bf Robertson-type uncertainty relation}
\begin{eqnarray*}
\Delta A_{\rho} \Delta B_{\rho} \geq \frac{1}{2}\mbox{tr}
(\rho[\hat{A},\hat{B}])|
%,
%where
\end{eqnarray*}
where
$\Delta A_{\rho}$ $(\Delta B_{\rho})$ represents standard deviation of the 
outcome of the corresponding observables. 
However, 
to characterize randomness of measurement outcomes, the standard 
deviation is often insufficient. 
The standard deviation depends on how to assign a value of 
measurement outcome to each event. For instance, let us imagine 
an observable which takes $0$, $1$ and $2$ as its value of 
measurement outcome. If a state gives 
an outcome $0$ or $1$ with 
probability $1/2$, its standard deviation is $1/2$. On the other hand, 
if we shuffle the values of outcome so that 
the new observable takes an outcome $0$ or $2$ 
with probability $1/2$, its standard 
deviation becomes $1$. 
In addition, the above Robertson-type formulation cannot deal with 
the most general type of measurement, positive operator 
value measure (POVM) measurement.
%, where a POVM $X=\{X_x\}$ is 
%a family of positive operators satisfying $\sum_x X_x ={\bf 1}$.
%\par
The entropic uncertainty relation can cover this type of 
measurement and is of advantage to its application.
It has the following form:
%\\
%{\bf Entropic Uncertainty Relation}
\begin{eqnarray*}
H(A|\rho)+H(B|\rho)\geq -2 \log \max_{a,b}\Vert A_a^{1/2} B_b^{1/2}\Vert,
\end{eqnarray*}
where $A:=\{A_a\}$ and $B:=\{B_b\}$ are POVMs and 
$H(A|\rho)$ $(H(B|\rho))$ represents Shannon 
entropy of the probability distribution 
of the measurement outcome of $A$ $(B)$ in a state $\rho$, 
i.e., $H(A|\rho)=-\sum_{a\in {\Omega}}\mbox{tr}(\rho A_a)
\log \mbox{tr}(\rho A_a)$.
This type of 
uncertainty relation
 was first proposed by Deutsch\cite{Deutsch} and 
 was improved by Maassen and Uffink\cite{MU}. 
 The above general form for POVMs was 
obtained by Krishna and Parthasarathy\cite{KP}. 
%%%%%%%%%%%%%%%%%%%%%%%%%%%%%%%%%%%%%%
\section{Remote Ensemble Preparation}\label{rep}
In this section we explain a way to prepare 
an ensemble of quantum states on a remotely 
located quantum system by using 
predistributed entangled state.
It %is called HJW theorem and 
plays 
an essential role 
to prove impossibility of 
the bit commitment.
It has been used to translate the 
BB84 quantum key distribution into 
E91 quantum key distribution.  
The theorem was first proved by 
Hughston, Jozsa and Wootters\cite{HJW}, and generalized by 
Halvorson\cite{Halvorson} for the most general quantum system including 
infinite systems. We, in this paper, treat only finite quantum systems
that are described by finite dimensional Hilbert spaces.  
Suppose there exist two characters: Alice and Bob. 
Each of them has a quantum system. 
The system possessed by Alice (Bob) is 
called as system $A$ (system $B$). 
Alice (Bob) can manipulate only the system $A$ (system $B$).
%Let us consider two remote systems: system $A$ belonging to 
%Alice and system $B$ to Bob. 
The system $A$ (system $B$) is described by 
a Hilbert space ${\cal H}_A$
(${\cal H}_B$). We assume that they have an identical finite dimension, 
${\cal H}_A \simeq {\cal H}_B \simeq {\bf C}^N$. 
We consider a method to 
prepare an ensemble of states on the system $B$ by 
Alice's operation on a predistributed entangled state
$|\Phi\rangle$.
A normalized vector of the composite system, 
$|\Phi\rangle \in {\cal H}_A \otimes {\cal H}_B$, can be 
written as, thanks to Schmidt decomposition theorem, 
\begin{eqnarray*}
|\Phi\rangle =\sum_k \sqrt{\lambda_k}|e_k^A \rangle \otimes |e_k^B\rangle,
\end{eqnarray*}  
where $\{|e^A_k\rangle \}$ ($\{|e^B_k\rangle\}$)
is an orthonormal basis of ${\cal H}_A$ (${\cal H}_B$).
We hereafter fix a normalized vector $|\Phi\rangle$ and 
its corresponding basis. ${\cal H}_A$ and ${\cal H}_B$ are 
identified with respect to these basis. We write its reduced 
state on each system as, 
\begin{eqnarray*}
\rho^A&=& \sum_k \lambda_k |e_k^A \rangle \langle e_k^A|
\\
\rho^B&=& \sum_k \lambda_k |e_k^B\rangle \langle e_k^B|.
\end{eqnarray*}
When we identify these two Hilbert space, 
we simply write them as $\rho(\equiv \rho^A \equiv \rho^B)$.
Suppose that the state $\rho^A$ can be decomposed into 
a mixture of the states as $\rho^A=\sum_i p_i \rho^A_i$, 
where $\rho^A_i$ is a state of the 
system $A$ for each $i$ and $\{p_i\}$ satisfies 
$\sum_i p_i=1$ and $p_i \geq 0$. Hereafter, for simplicity, 
we assume $\mbox{rank}\rho=N$. 
In the following, we consider a measurement by Alice 
that prepares the state $\rho_i$ with the probability $p_i$ 
on the system $B$ attached to Bob.
%%%%%%%%%%%%%%%%
We define transpose operation with respect to 
the basis $\{|e_k^A\rangle\}$. 
Since the transpose operation $A \mapsto \ ^tA$ preserves 
the positivity of the operator, 
a family of operators, 
\begin{eqnarray}
F[\{p_j,\rho_j\}]
%:=\{F[\{p_j,\rho_j\}]_i\}:=\{ p_i \ ^t \rho^{-1/2} \ ^t 
%\rho_i \ ^t \rho^{-1/2}\}$ 
:=\{F[\{p_j,\rho_j\}]_i\}:=\{ p_i \rho^{-1/2} \ ^t 
\rho_i \rho^{-1/2}\}
\label{Fdef}
\end{eqnarray} 
forms a POVM. 
%, where 
%the transpose operation $A\mapsto \ ^tA$ is defined 
%with respect to the basis $|e\
%(Note that $\ ^t(A=(\ ^tA)^{-1/2}$ holds for 
%an arbitrary positive operator $A$.)
Let us take the state $|\Phi\rangle$ and consider 
an a-posteriori state with respect to the 
POVM $F[\{p_j,\rho_j\}]$. 
A probability to obtain an outcome $i$ is 
calculated as 
\begin{eqnarray*}
\langle \Phi|F[\{p_j,\rho_j\}]|\Phi\rangle
&=&\sum_k \lambda_k \langle e^A_k|F[\{p_j,\rho_j\}]_i|e^A_k\rangle
\\
&=&
\sum_{k} \lambda_k p_i \langle e^A_k|\rho^{-1/2}\rho_i\rho^{-1/2}|e^A_k\rangle
\\
&=& 
p_i \mbox{tr}(\rho \rho^{-1/2}\rho_i \rho^{-1/2})
=p_i.
\end{eqnarray*}
%The aposteriori state is calculated as, 
Since Alice does not 
make any operation on the system $B$, the a-posteriori state of the system $B$ 
for the outcome $i$ is calculated as
$\langle \Phi|F[\{p_j,\rho_j\}]_i A|\Phi\rangle /p_i$.
Since for each operator $A$ on ${\cal H}_B$, 
\begin{eqnarray*}
%&&
\langle \Phi|F[\{p_j,\rho_j\}]_i A|\Phi\rangle
%\\
&=&
\sum_{k,l} \sqrt{\lambda_k \lambda_l}
\langle e_k |\rho^{-1/2}\ ^t \rho_i \rho^{-1/2}|e_l\rangle
\langle e_k|A|e_l \rangle
\\
&=&
%&&
\sum_{k,l} \sqrt{\lambda_k \lambda_l}
\langle e_l |\rho^{-1/2}\rho_i \rho^{-1/2}|e_k\rangle
\langle e_k|A|e_l \rangle
\\
&=&
%=
p_i \mbox{tr}(\rho_i A) 
\end{eqnarray*}
holds, where we used a relation $\ ^t \rho =\rho$,
the a-posteriori state of the system $B$ is $\rho_i$. 
%Thus we constructed the measurement by Alice 
%that prepares the wanted ensemble to Bob's side.
We thus proved the following theorem. 
%%%%%%%%%%%%%%%%%
\begin{theorem}
Suppose that there exist Hilbert spaces ${\cal H}_A$, ${\cal H}_B$ 
 and a normalized vector $|\Phi\rangle
\in {\cal H}_A \otimes {\cal H}_B$. 
Assume that the reduced density operator $\rho:=
\mbox{tr}_{{\cal H}_B}(|\Phi\rangle \langle \Phi|)$ 
can be decomposed into a mixture of states as,
$\rho=\sum_i p_i \rho_i$. 
%A POVM $F[\{p_j,\rho_j\}]:=\{F[\{p_j,\rho_j\}]_i\}$ on 
%${\cal H}_A$ is defined $F[\{p_j,\rho_j\}]_i
%:=p_i \rho^{-1/2} \ ^t \rho_i \rho^{-1/2}$, where 
%the transpose operation $A\maspto \ ^t A$ is 
%defined with respect to the
%If one makes a measurement of ${\cal H}_A$ 
%with respect to a POVM $F[\{p_i,\rho_i\}]$ 
There exists a POVM $F[\{p_j,\rho_j\}]=\{F[\{p_j,\rho_j\}]_i$ 
on ${\cal H}_A$ that prepares the ensemble $\{p_i,\rho_i\}$ 
on ${\cal H}_B$.  That is, the probability to obtain an
outcome $i$ is $p_i$, and the a-posteriori state 
of ${\cal H}_B$ then is $\rho_i$.
\end{theorem}
%%%%%%%%%%%%%%%%%%%%%%%%%%%%%%%%%%%%%%%%%%%%%%%%%%%
\section{Information-Disturbance theorem}\label{infodis}
In this section, we derive two types of Information-Disturbance 
theorem. Both treat a cryptographic setting. The first one
relates information gain by Eve with information gain by Bob.
The second one relates information gain by Eve with 
randomness of error contained in Bob's outcome.
%%%%%%%%%%%  
\subsection{Information v.s. Information}
We deal with a quantum cryptographic setting. It is a 
simplified version of the BB84 protocol. Three 
characters: Alice, Bob, and Eve, play their roles. 
Alice has a quantum system described by an $N$-dimensional 
Hilbert space, ${\cal H}_A$. She prepares a state $\rho$ of this 
system in one of the two different methods:
%A method is that 
(a) 
she prepares a state $\rho_i$ with probability $p_i$ 
for each $i$, %and another is that 
or 
(b)
she 
prepares a state $\sigma_l$ with probability $q_l$ for each $l$. 
To assure that both procedures actually give the state $\rho$, we 
impose a condition, 
$\rho=\sum_i p_i \rho_i=\sum_l q_l \sigma_l$.
We write $X$ ($Y$) a random variable whose value takes $i$ ($l$) 
with 
the probability $p_i$ ($q_l$). 
The preparation can be regarded as 
encoding $X$ or $Y$ to the state $\rho$.
The full protocol runs as follows:
\begin{itemize}
\item[(i)] Alice chooses one of the two methods, (a) or (b), to prepare 
the state $\rho$.
\item[(ii)] Alice prepares the state $\rho$ according to 
her choice on the method. That is, Alice encodes $X$ or $Y$ to the state. 
\item[(iii)] Alice sends the system to Bob. 
\item[(iv)] After confirming that Bob has actually 
received the system, Alice publishes the method ((a) or (b)) 
which she employed to prepare the state $\rho$.
\item[(v)]
Bob makes a measurement on his received system to 
extract the encoded information.
We write Hilbert space of the received system as ${\cal H}_B$ 
instead of ${\cal H}_A$ for convenience.
\end{itemize}
Note that even if there is no eavesdropper between Alice and Bob, 
Bob does not obtain in general the full information Alice has encoded. 
That is, the encoding employed by Alice may be 
ambiguous one. $\rho_i$ and $\rho_j$ for $i\neq j$ may not be 
distinguishable perfectly.
In the following we will see that 
Eve's eavesdropping in addition makes Bob's information gain less. 
Let us see what Eve can do.
Eve who wants to obtain information encoded by Alice 
can make her own apparatus %${\cal H}_E$ 
interact with 
${\cal H}_A$ %\simeq {\cal H}_B$ 
when it is sent to Bob. 
She may keep the apparatus and only after knowing 
Alice's announcement, may make a measurement on it
to obtain a classical output. 
Denote ${\cal H}_E$ the Hilbert space of Eve's apparatus.
%Without loss of generality,
Eve's operation %with the procedure (iii) 
is described by a unitary operator 
%$U$,
%\begin{eqnarray}
$
U:{\cal H}_B \otimes {\cal H}_E \to {\cal H}_B \otimes {\cal H}_E. 
$
A state of the apparatus before the interaction is 
written as $|\Omega\rangle \in {\cal H}_E$. 
Without loss of generality, we can assume it 
as vector state. 
 Eve's attack is determined by the triplet,
 $({\cal H}_E, |\Omega \rangle, U)$.
After Alice's announcement, 
Eve tries to make an optimal
 measurement $Z$, a POVM, 
 on her apparatus to 
 extract the encoded information.
 %,
 %where a POVM is a family of the positive operators $\{X_{\alpha} 
 %\}_{\alpha}$ 
 %satisfying $\sum_{\alpha} X_{\alpha}={\bf 1}$.
% We write again $Z$ the random variable representing 
% the outcome with respect to the POVM $Z$. 
% by using the knowledge of the 
% basis.
What we are interested in is the trade-off between
the information gain by Bob and one by Eve. 
%%%%%%%%%%%%%%%%%%%%%%%
Let us suppose a fixed Eve's attack $({\cal H}_E,
|\Omega\rangle, U)$. We define $I(X:B)$ as optimal
information gain by Bob on random variable $X$. 
That is, if Alice has encoded $X$ to the quantum state
and Eve employs an attack $({\cal H}_E, 
|\Omega\rangle, U)$, Bob's optimal  
information gain on $X$ is $I(X:B)$. 
%when Alice has encoded $X$ to the quantum state $\rho$. 
In the same manner, $I(Y:B)$ is defined as 
information gain by Bob on random variable $Y$. 
$I(Y:E)$, on the other hand, is defined 
as optimal information gain by Eve on $Y$ if 
Alice has encoded $Y$ to the quantum state $\rho$ 
and Eve employs the attack $({\cal H}_E, |\Omega\rangle,
U)$.
$I(X:E)$ is defined as optimal information gain by Eve on 
$X$. 
\begin{theorem}
For a fixed Eve's attack $({\cal H}_E, |\Omega\rangle, U)$, 
the following inequalities hold:
 \begin{eqnarray*}
%&&
I(X:B)+I(Y:E) &\leq& H(X)+H(Y) %\\
%&&
+2\log  \max_{i,k} \Vert F[\{p_j,\rho_j\}]_i^{1/2}
F[\{q_l,\sigma_l\}]_k^{1/2} \Vert,
\\
%&&
I(X:E)+I(Y:B) &\leq& H(X)+H(Y) %\\
%&&
+2\log  \max_{i,k} \Vert F[\{p_j,\rho_j\}]_i^{1/2}
F[\{q_l,\sigma_l\}]_k^{1/2} \Vert,
\end{eqnarray*}
where POVMs, $F[\{p_j,\rho_j\}]$ and $F[\{q_l,\sigma_l\}]$ 
are defined by (\ref{Fdef}).
\end{theorem}
{\bf Proof:}
\\
%%%%%%%%%%%%%%%%%%%%%%
To calculate Bob's and Eve's information gain, we 
construct an appropriate probability distribution. 
We apply the remote ensemble preparation technique.
Suppose that $\rho$ can be diagonalized 
as $\rho=\sum_k \lambda_k |e_k^A \rangle\langle e_k^A|$
and thus 
$\{|e_k^A\}$ forms a basis of ${\cal H}_A$.  
We introduce $\{|e_k^B\rangle\}$, a basis of ${\cal H}_B$ and 
use $\{|e_k^A\rangle\}$ and $\{|e_k^B\rangle\}$ to 
identify both Hilbert spaces.  
Let us introduce a virtual entangled state on 
${\cal H}_A \otimes {\cal H}_B$. 
A normalized vector $|\Phi\rangle \in {\cal H}_A \otimes {\cal H}_B$ 
is defined as, 
\begin{eqnarray*}
|\Phi\rangle:=\sum_k \sqrt{\lambda_k} |e_k^A \rangle \otimes |e_k^B
\rangle.
\end{eqnarray*}
%As we have explained, this state can produce a 
%probability distribution for Alice and Bob in case of inexistence of Eve. 
As we have explained, this state can be used for 
the remote ensemble preparation in case 
of existence of Eve.
In fact, if Alice operates a POVM $F[\{p_i,\rho_i\}]$ 
$(F[\{q_l,\sigma_l\}])$ 
on this state, Bob obtains a state $\rho_i$ $(\sigma_l)$ with 
probability $p_i$ $(q_l)$. 
The effect Eve gives on it can be included by 
defining a new state, 
\begin{eqnarray*}
|\Psi\rangle:= ({\bf 1}\otimes U)|\Phi\rangle \otimes |\Omega\rangle. 
\end{eqnarray*}
If Alice applies a POVM $F[\{p_i,\rho_i\}]$ $(F[\{q_l,\sigma_l\}])$ on 
this state, she obtains an outcome $i$ $(l)$ 
with probability $p_i$ $(q_l)$
and the state of Bob and Eve then is 
$U(\rho_i \otimes |\Omega\rangle \langle \Omega|)U^*$
$(U(\sigma_l \otimes |\Omega\rangle \langle \Omega|)U^*)$.
Let us consider arbitrary POVMs $\tilde{B}:=\{\tilde{B}_b\}$ of Bob's and 
$Z:=\{Z_z\}$ of Eve's. We write the random variable 
representing the outcome of Bob's (Eve's) measurement also as
$\tilde{B}$ ($Z$). 
A-posteriori state $\rho_{\tilde{B}=b,Z=z}$ 
of ${\cal H}_A$
with respect to these POVMs is written as 
\begin{eqnarray*}
%\omega(\cdot|B=b,Z=z)
%:=\frac{\omega(\cdot \tilde{B}_b Z_z)}{\omega(\tilde{B}_b Z_z)}.
\mbox{tr}(\rho_{\tilde{B}=b,Z=z}A)
:=\frac{\mbox{tr}(\rho A \tilde{B}_b Z_z)}
{\mbox{tr}(\rho \tilde{B}_b Z_z)}
\end{eqnarray*}
for an arbitrary operator $A$ on ${\cal H}_A$.
We apply the entropic uncertainty relation to 
this state. The observables to be concerned are 
POVMs: $F[\{p_j,\rho_j\}]$ and $F[\{q_l,\sigma_l\}]$. 
We obtain, 
\begin{eqnarray*}
%&&
H(X|\tilde{B}=b,Z=z)+H(Y|\tilde{B}=b,Z=z)
%\\
\geq
-2 \log \max_{i,k} \Vert F[\{p_j,\rho_j\}]_i^{1/2}
F[\{q_l,\sigma_l\}]_k^{1/2}\Vert.
\end{eqnarray*}
Subtracting $H(X)+H(Y)$ from both sides and 
summing them up with $\langle \Phi|\tilde{B}_b Z_z|\Phi\rangle$, 
we obtain, 
\begin{eqnarray*}
%&&
I(X:\tilde{B},Z)+I(Y:\tilde{B},Z)&\leq& H(X)+H(Y)
\\
&&
+2 \log \max_{i,k} \Vert F[\{p_j,\rho_j\}]_i^{1/2}
F[\{q_l,\sigma_l\}]_k^{1/2}\Vert.
\end{eqnarray*}
Using $I(X:\tilde{B})\leq I(X:\tilde{B},Z)$
 and $I(Y:Z)\leq I(Y:\tilde{B},Z)$,
or $I(X:Z)\leq I(X:\tilde{B},Z)$ and $I(Y:\tilde{B})\leq 
I(Y:\tilde{B},Z)$
we obtain,
\begin{eqnarray*}
%&&
I(X:\tilde{B})+I(Y:Z) &\leq& H(X)+H(Y) %\\
%&&
+2\log  \max_{i,k} \Vert F[\{p_j,\rho_j\}]_i^{1/2}
F[\{q_l,\sigma_l\}]_k^{1/2} \Vert,
\\
%&&
I(X:Z)+I(Y:\tilde{B}) &\leq& H(X)+H(Y)
%\\
%&&
+2\log  \max_{i,k} \Vert F[\{p_j,\rho_j\}]_i^{1/2}
F[\{q_l,\sigma_l\}]_k^{1/2} \Vert.
\end{eqnarray*}
Since the POVMs $\tilde{B}$ and $Z$ are arbitrary, 
we can take the optimal one for both. 
\hfill Q.E.D.
\par
This theorem gives nontrivial bounds if POVMs $F[\{p_j,\rho_j\}]$
and $F[\{q_l,\sigma_l\}]$ do not commute with each other. 
That is, when Eve employs an operation that should yield herself to 
obtain large information if the encoded random variable was 
$Y$ $(X)$, Bob cannot obtain large information on $X$ $(Y)$ that was 
actually employed by Alice. 
\par 
Let us consider the simplest example.
The system consists of $N$-qubits. $\rho$ is the maximally mixed state, 
$\rho=\frac{{\bf 1}}{2^N}$. Each bit has two natural basis corresponding 
to the eigenvectors of $\sigma_z$ and $\sigma_x$. 
Let $b$ be an element of $\{z,x\}^N$. $b$ naturally determines 
a basis of $N$-qubit and an observable $X[b]$ that is diagonalized 
by this basis. We write $\overline{b}$, the conjugate basis of $b$. 
It is defined by exchange all $z$ $(x)$ to $x$ $(z)$. 
We write its corresponding observable as $X[\overline{b}]$. 
In this situation, we obtain, 
\begin{eqnarray*}
I(X[b]:B)+I(X[\overline{b}]:E)\leq N.
\end{eqnarray*} 
%\section{Information vs. Randomness of Error}
%We, in the previous paper, derived a relation between 
%inforation gain by Eve and randomness of error contained in Bob's data.
%We here rederive the result using the remote ensemble preparation 
%technique and the entropic uncertainty relation. 
%%%%%%%%%%%%%%%%%%%%%%%%%%%%%%%%%
\subsection{Information v.s. Randomness of Error}
In \cite{della} we derived a theorem \cite{Hayashi} that relates 
information gain by Eve and randomness of error contained 
in Bob's data. Its derivation, however, relied upon 
symmetrization technique and Holevo bound, and was 
complicated. We here give another simple proof of the theorem 
by the remote ensemble preparation technique and 
the entropic uncertainty relation.
% Our new derivation 
%enables us to slightly generalize the theorem. 
\par
Let us first begin with the setting. It is 
a special case of the above general one. 
Let us consider two pairs of orthogonal states,
$\{|0\rangle, |1\rangle\}$ and its {\it conjugate} 
$
\{|\overline{0}\rangle,|\overline{1}\rangle\}$
in ${\bf C}^2$.
%In contrast with \cite{della}, 
%they are not necesarily mutually unbiased with 
%each other. 
They are assumed mutually unbiased and thus 
\begin{eqnarray*}
|\langle i|\overline{j}\rangle|^2
=\frac{1}{2}
\end{eqnarray*}
holds for each $i,j=0,1$. 
Alice has $N$-qubits described by a Hilbert space
${\cal H}_A={\bf C}^2 \otimes {\bf C}^2 \otimes \cdots
\otimes {\bf C}^2$ ($N$ times). 
For each $i=i_1 i_2 \cdots i_N \in \{0,1\}^N$,
we write $|i\rangle=|i_1\rangle \otimes |i_2\rangle 
\otimes \cdots \otimes |i_N\rangle$ 
and $|\overline{i}\rangle :=
|\overline{i_1}\rangle \otimes |\overline{i_2} \rangle \otimes 
\cdots \otimes |\overline{i_N}\rangle$. 
She prepares a 
maximally mixed state $\rho=\frac{{\bf 1}}{2^N}$ 
of this system in one of the two different methods: 
$(a)$ she prepares a state $|i\rangle \langle i|$ 
with probability $\frac{1}{2^N}$ for each $i\in \{0,1\}^N$, 
or 
$(b)$ she prepares a state $|\overline{j}\rangle
\langle \overline{j}|$ with probability 
$\frac{1}{2^N}$ for each $j\in \{0,1\}^N$. 
%We introduce a quantity expressing their biasedness as,
%\begin{eqnarray*}
%p:=\max_{i.j=0,1}|\langle i|\overline{j}\rangle |.
%\end{eqnarray*}
%It takes a value between 
%$1/2$ and $1$, and takes $1/2$ if they are mually unbiased with each other. 
%\begin{eqnarray*}
%\langle i| \overline{k} \rangle =\sqrt{\frac{1}{2}}(-1)^{ik}
%\end{eqnarray*}
%holds for each pair of $i,k \in \{0,1\}$.
We write a random variable $A$ which takes value $i\in \{0,1\}^N$
with probability $\frac{1}{2^N}$. Alice encodes this 
random variable to quantum state $\rho$ by one of the 
methods $(a)$ or $(b)$.
\begin{itemize}
\item[(i)]
Alice first selects $(a)$ or $(b)$ which is used to 
encode a random number. 
\item[(ii)]
Alice encodes the random variable $A$ to
 the state $\rho=\frac{{\bf 1}}{2^N}$
according to her choice on the method. 
That is, if she has chosen $(a)$, Alice 
prepares $|i\rangle\langle i|$ with probability 
$\frac{1}{2^N}$. On the other hand, if her choice was 
$(b)$, she prepares $|\overline{j}\rangle \langle 
\overline{j}|$ with probability $\frac{1}{2^N}$.   
%Alice next randomly generates an $N$-bits sequence $i\in \{0,1\}^N$ with 
%probability $p(i)=\frac{1}{2^N}$. 
%We write $A$ a random variable representing this $N$-bits sequence.
%Typically $p(i)$ is equally waited, $p(i)=\frac{1}{2^N}$.
%\item[(iii)]
%Alice encodes this information on $N$-qubits and sends them to Bob.
%For instance, suppose that Alice selects $b$ and generates a 
%sequence $i=i_1 i_2\cdots i_N$, she sends the corresponding state $|i\rangle
%=|i_1\rangle \otimes |i_2 \otimes \cdots \otimes |i_N\rangle \in {\bf C}^2 
%\otimes \cdots \otimes {\bf C}^2=:{\cal H}_A\simeq {\cal H}_B$ to Bob.
%If the conjugate
% basis $\overline{b}$ and a sequence $i=i_1 i_2 \cdots i_N$ are 
%chosen, 
%the state sent to Bob is $|\overline{i}\rangle=
%|\overline{i_1}\rangle \otimes |\overline{i_2}\rangle 
%\otimes \cdots \otimes |\overline{i_N}\rangle
%\in {\cal H}_A$.
\item[(iii)]
Alice sends the system to Bob. 
\item[(iv)]
Alice, after confirming that Bob actually 
has received $N$-qubits, informs him 
of the method ($(a)$ or $(b)$) she used. 
\item[(v)]
Bob makes a measurement with respect to 
the basis and obtains an outcome. Let us 
write $B$ the random variable representing 
this outcome. If there is no eavesdropper, 
$A=B$ naturally follows.
\end{itemize}
Eve wants to obtain the information of the random variable $A$.
For the purpose, Eve prepares an apparatus and makes it interact with 
the $N$-qubits sent to Bob by Alice.  
Denote ${\cal H}_E$ the Hilbert space of Eve's apparatus.
%Without loss of generality,
Eve's operation is described by a unitary operator 
%$U$,
%\begin{eqnarray}
$
U:{\cal H}_B \otimes {\cal H}_E \to {\cal H}_B \otimes {\cal H}_E. 
$
A state of the apparatus before the interaction is 
written $|\Omega\rangle \in {\cal H}_E$.
 Thus Eve's attack is determined by the triplet,
 $({\cal H}_E, |\Omega \rangle, U)$.
After the publication of the basis, 
Eve tries to make an optimal
 measurement $Z=\{Z_z\}$, a POVM (positive operator valued measure), 
 on her apparatus to 
 extract the information of $A$. 
 %,
 %where a POVM is a family of the positive operators $\{X_{\alpha} 
 %\}_{\alpha}$ 
 %satisfying $\sum_{\alpha} X_{\alpha}={\bf 1}$.
% We write $E[Z]$, the random variable representing 
% the outcome with respect to the POVM $Z$.

% by using the knowledge of the 
% basis.
What we are interested in is the trade-off between
the information gain by Eve and the errors contained in 
Bob's outcome.
Let us suppose a fixed Eve's attack. 
We define $I(A:E|a)$ as Eve's optimal information gain 
on $A$ if Alice has chosen the method $(a)$ for encoding. 
$I(A:E|b)$ is defined as Eve's optimal information gain on $A$ 
if Alice has chosen the method $(b)$. 
%%%%%%%%%%%%%%%%%%%%%%%%%
We can show the following theorem.
 %%%---------------------------------------------------------
 \begin{theorem}
Information gain by Eve inevitably makes
Bob's data in another basis random.
More precisely, for the fixed 
Eve's attack $({\cal H}_E, |\Omega\rangle, U)$, 
%$U$ and $X$, 
the following inequality holds:
%for the probability distribution (\ref{equiv}):
\begin{eqnarray*}
I(A:E|a)%-N \log (2p)
\leq H(A\oplus B|b),
\label{Ours}
\end{eqnarray*}
%where $I(\cdot:\cdot)$ represents the 
%mutual information and $H(\cdot)$ is the Shannon entropy.
where $H(A\oplus B|a)$ is the Shannon entropy of the error 
contained in Bob's outcome when Alice has chosen the method 
$(a)$ for encoding. 
\end{theorem}
{\bf Proof:}
For the proof, we employ the remote 
ensemble preparation technique.
Since the state prepared by Alice is 
maximally mixed state, the relevant entangled state of 
${\cal H}_A \otimes {\cal H}_B$ is 
maximally entangled state:
$|\Phi \rangle :=
\frac{1}{2^N} \sum_i |i\rangle \otimes |i\rangle$. 
The effect of Eve's attack can be included by 
defining a new state, 
\begin{eqnarray*}
|\Psi\rangle := ({\bf 1} \otimes U) |\Phi\rangle \otimes |\Omega\rangle. 
\end{eqnarray*}
Now suppose that Eve employed a POVM
$Z:=\{Z_{z} \}$ and obtained 
a value $z$. 
We write $\rho_{z}$, a-posteriori state on ${\cal H}_A\otimes {\cal H}_B$.
To this state, we apply the entropic uncertainty relation.
To introduce relevant POVMs, we fix a basis 
to define transpose operation as $\{|i\rangle\}$.
It is convenient to introduce a 
new basis $\{|\underline{i}\rangle \}$ as
 $|\underline{i}\rangle:=
 \sum_j |j\rangle \langle \overline{i}|j\rangle$. 
 In fact, the transposition of $|\overline{i}\rangle
 \langle \overline{i}|
 =\sum_{j,k}|j\rangle \langle j|\overline{i}\rangle
 \langle \overline{i}|k\rangle \langle k|$ 
 with respect to the basis $\{|i\rangle \}$
 can be simply written as,
 $|\underline{i}\rangle \langle \underline{i}|$. 
 Let us define observables $F_A$ and $F_{\underline{A}}$ on ${\cal H}_A$ as
 $F_A:=\sum_{i\in\{0,1\}^N} i|i\rangle \langle i|$
 and $F_{\underline{A}}
 :=\sum_{i\in \{0,1\}^N} i|\underline{i}\rangle \langle \underline{i}|$.
 Let us define observable $G_{\overline{B}}$ 
 on ${\cal H}_B$ as
% $G_B:=\sum_{i\in \{0,1\}^N} i|i\rangle \langle i|$ 
% and 
 $G_{\overline{B}}:=\sum_j j |\overline{j}\rangle \langle \overline{j}|$.
Observables to be treated are
$F_{\overline{A}}\oplus G_{\overline{B}}=\sum_l l E_l$
that gives probability distribution for $A\oplus B$ in $(b)$
and $F_{A}\otimes {\bf 1}=\sum_j jP_j$ 
that gives probability distribution for $A$ in $(a)$.
The following inequality holds:
\begin{eqnarray}
%&&
H(F_{\overline{A}}\oplus G_{\overline{B}}|Z=z)
+H(F_{A}\otimes {\bf 1}|Z=z)
%\nonumber
%\\
\geq -2 \log\left( \max_{l,j} \Vert E_l P_j \Vert \right). 
\label{entalpha}
\end{eqnarray}
Thus we must estimate 
$\Vert E_k P_j \Vert$.
From 
\begin{eqnarray*}
F_{\overline{A}}\oplus G_{\overline{B}}=
\sum_{l,i} l |\underline{i}\rangle 
\langle \underline{i}|\otimes |\overline{i\oplus l}
\rangle \langle \overline{i \oplus l}|,
\end{eqnarray*}
we obtain
$E_l=\sum_i |\underline{i}\rangle \langle
\underline{i}|\otimes |\overline{i\oplus l}\rangle
\langle \overline{i\oplus l}|$.
Therefore
\begin{eqnarray*}
E_l P_j= \sum_i |\underline{i}\rangle
\langle \underline{i}|j\rangle \langle j|
\otimes |\overline{i\oplus l}\rangle \langle \overline{i\oplus l}|
\end{eqnarray*}
holds. To estimate the norm of this operator,
we introduce a normalized vector $|\Phi\rangle :=
\sum_{k u}\alpha_{k u}|k\rangle \otimes |\overline{u}\rangle$
and apply $E_l P_j$ on it.
\begin{eqnarray*}
E_l P_j|\Phi\rangle
=\sum_{i}
\alpha_{j i \oplus l}|\underline{i}\rangle
\otimes |\overline{i\oplus l}\rangle\langle
\underline{i}|j\rangle 
\end{eqnarray*}
gives 
\begin{eqnarray*}
\Vert E_l P_j|\Phi\rangle\Vert^2
&=&
\sum_i |\alpha_{j i\oplus l}|^2 |\langle \underline{i}|
j\rangle |^2 \\
&\leq&
\max_{i } |\langle \underline{i}|j\rangle|^2
\sum_i |\alpha_{j i \oplus l}|^2
\\
&\leq&
\max_{i}|\langle \underline{i}|j\rangle|^2.
\end{eqnarray*}
Thanks to $|\langle  \underline{i}|j\rangle|
=|\langle \overline{i}|j\rangle|$,
we obtain
\begin{eqnarray*}
\max_{l j}\Vert E_l P_j\Vert
\leq (\frac{1}{2})^{N/2}.
\end{eqnarray*}
Application of this inequality to (\ref{entalpha})
leads us 
 \begin{eqnarray*}
%H(A \oplus B|E[X]=\alpha,b=1)
%+H(A|E[X]=\alpha,b=0)
H(F_{\overline{A}}\oplus G_{\overline{B}}|Z=z)
+H(F_{A}\otimes {\bf 1}|Z=z)
\geq N 
\end{eqnarray*}
Taking an average with respect to $z$ 
%in $p(\alpha)$ (probability that Eve obtains an outcome $\alpha$) 
and adding $N$ to both sides, we obtain
\begin{eqnarray*}
I(A:Z|a)\leq H(A\oplus B|b).
\end{eqnarray*}
Since the POVM $Z$ is arbitrary, we obtain the theorem. 
%In the original variable, the final form of the inequality becomes
%The case when Alice has chosen $b=1$ runs in the same manner.
%We finally obtain,
%\begin{eqnarray*}
%I(A:E[X]|b)-N \log(2p)\leq H(A\oplus B|\overline{b}).
%\end{eqnarray*}
\hfill {\bf Q.E.D.}
\par
%The unbiased case, corresponding to $p=1/2$, coincides with our 
%previous result\cite{della} which was proven by the different method.
%%%%%%%%%%%%%%%%%%%%%%%%%%%%%%%%%%%%
\section{Summary}
In conclusion, we derived a generalization of 
the information-disturbance theorm. Our generalized 
theorem can treat a general source (a pair of ensembles 
that give the same state) and relate Eve's information gain 
for an ensemble with Bob's information gain
for another ensemble. The result is a direct 
consequence of the entropic uncertainty relation. 
\section*{Acknowledgment}
The authors would like to thank 
Kentaro Imafuku for helpful discussions.


\begin{thebibliography}{1}

\bibitem{BB84}
C.~H.~Bennett and G.~Brassard. 
\newblock
Quantum Cryptography: Public Key Distribution and Coin Tossing.
\newblock In {\em Proc. of 
IEEE Int. Conf. on Computers, Systems and Signal Processing}, 
pages 175--179, 1984.
\bibitem{Mayers}
D.~Mayers.
\newblock
 Quantum key distribution and string oblivious transfer in noisy 
channel. 
\newblock
In {\em Advances in cryptology - CRYPTO'96}, 
\newblock LNCS 1109, pages 343--357, 1996.
\bibitem{HKL}
H-K.~Lo and H-F.~Chau.
%\newblock 
%Unconditional security of quantum key distribution over 
%arbitrary long distances. 
\newblock {\em Science}, 283, pages 2050--2056, 1999. 

\bibitem{Shor}P.~W.~Shor and J.~Preskill. 
%\newblock 
%Simple proof of security of the BB84 
%quantum key distribution protocol. 
\newblock {\em Phys.Rev.Lett.},
85, pages 441--444, 2000.
\bibitem{Biham}E.~Biham, M.~Boyer, P.~O.~Boykin, T.~Mor, and 
V.~Roychowdhury.
\newblock
A proof of the security of quantum key distribution.
\newblock
in {\em Proc. of the 32nd Annual ACM Symposium on Theory of Computing}, 
pages 715--724, 2000. :
E.~Biham, M.~Boyer, P.~O.~Boykin, T.~Mor, and 
V.~Roychowdhury.
%\newblock
%A proof of the security of quantum key distribution.
%\newblock
%Accepted for publication in {\em J. of Cryptology}, quant-ph/0511175.
%\bibitem{noclone}
%W.K.~Wootters, and W.H.~Zurek. 
%\newblock
%A Single Quantum Cannot be Cloned. 
%\newblock
%{\em Nature} 299, pages 802-803 1982.
\bibitem{FuchsPeres}
C.~A.~Fuchs and A.~Peres.
%\newblock
%Quantum State Disturbance vs. Information Gain: 
%Uncertainty Relations for Quantum Information,
\newblock {\em Phys.Rev.A}, 53(4), pages 2038--2045, 1996.
\bibitem{Fuchs}
C.~A.~Fuchs.
%\newblock
%Information Gain vs. State Disturbance in Quantum Theory,
\newblock {\em Fortschritte der Physik}, 46(4,5), pages 535--565, 1998.
\bibitem{Winter}
M.~Christandl and A.~Winter.
%\newblock
%Uncertainty, Monogamy, and Locking of Quantum Correlations,
\newblock
{\em IEEE Trans Inf Theory}, 51(9), pages 3159--3165, 2005.
\bibitem{Boykin}
P.~O.~Boykin and V.~P.~Roychowdhury.
%\newblock 
%Information vs. disturbance in dimension D.
\newblock
{
\em QIC: Quantum Information and Computation}, 5(5), pages 396--412, 2005. 
\bibitem{della}
T.~Miyadera and H.~Imai,
%\newblock
%Information-Disturbance Theorem for Unbiased Observables,
{\em Phys.Rev.A.} 73, pages 042317 2006.
\bibitem{NC}M.~A.~Nielsen, and I.~L.~Chuang,
{\it Quantum Computation and Quantum Information},  
Cambridge press. 2000.
%\bibitem{SCIS2006}T.~Miyadera and H.~Imai,
%Quantum Bounded Storage Model and 
%Uncertainty Relations,
%SCIS2006.
\bibitem{Deutsch}D.~Deutsch, {\em Phys.Rev.Lett.} 50,631 (1983).
\bibitem{MU}H.~Maassen, and J.~Uffink, 
{\em Phys. Rev. Lett.}, 60, pages 1103 (1998).
\bibitem{KP}M.~Krishna and K.~R.~Parthasarathy,
%An entropic uncertainty principle for quantum measurements,
{\em Sankhya, Series A,} 64(3), 842 (2002). 
\bibitem{HJW}
L.~Hughston, R.~Jozsa, and W.~Wootters, 
{\em Phys. Lett. A.} 183 pages 14 (1993).
\bibitem{Halvorson}
H.~Halvorson, {\em J. Math. Phys.} 45, pages 4920 (2004).
%\bibitem{OzawaBayes}M.~Ozawa, Ann. Phys. {\bf 259},121 (1997).
\bibitem{Hayashi}
It was derived independently by M. Hayashi: 
M.~Hayashi, {\em Phys. Rev. A.}
74, pages 022307 2006.
\end{thebibliography}
\end{document}